\documentclass[10pt, conference, compsocconf]{IEEEtran}

\usepackage[all]{xy}
\usepackage{color,url}
\usepackage{subfigure}
\usepackage{bm,amsmath}
\usepackage{marvosym}
\usepackage{graphicx}
\usepackage{wrapfig}
\usepackage{epsf}
\usepackage{times,mathptm}
\usepackage{url}
\usepackage{epstopdf}
\usepackage{epsfig}

\usepackage{algorithmic}
\usepackage{algorithm2e}
\usepackage{graphicx}
\graphicspath{Figures/}

\begin{document}

\title{Pressure Fluctuations in Natural Gas Networks\\ caused by Gas-Electric Coupling}
%following Renewable- Generators:\\ Gas-Grid Coupling}

% author names and affiliations
% use a multiple column layout for up to two different
% affiliations

\author{
\IEEEauthorblockN{Misha Chertkov}
\IEEEauthorblockA{T-4 \& CNLS, LANL\\
Los Alamos, NM\\
chertkov@lanl.gov}
\and
\IEEEauthorblockN{Michael Fisher}
\IEEEauthorblockA{
T-4, LANL,
Los Alamos, NM\\
\& EECS, U of Michigan\\
Ann Arbor, MI\\ fishermw@umich.edu}
\and
\IEEEauthorblockN{Scott Backhaus}
\IEEEauthorblockA{MPA, LANL\\
Los Alamos, NM\\
backhaus@lanl.gov}
\and
\IEEEauthorblockN{Russell Bent}
\IEEEauthorblockA{DSA-4, LANL\\
Los Alamos, NM\\
rbent@lanl.gov}
\and
\IEEEauthorblockN{Sidhant Misra}
\IEEEauthorblockA{EECS, MIT\\
Cambridge, MA\\
sidhant@mit.edu}
}

% make the title area
\maketitle

\begin{abstract}
The development of hydraulic fracturing technology has dramatically increased the supply and lowered the cost of natural gas in the United States, driving an expansion of natural gas-fired generation capacity in several electrical interconnections. Gas-fired generators have the capability to ramp quickly and are often utilized by grid operators to balance intermittency caused by wind generation.  The time-varying output of these generators results in time-varying natural gas consumption rates that impact the pressure and line-pack of the gas network.
As gas system operators assume nearly constant gas consumption when estimating pipeline transfer capacity and for planning operations, such fluctuations are a source of risk to their system. Here, we develop a new method to assess this risk. We consider a model of gas networks with consumption modeled through two components: forecasted consumption and small spatio-temporarily varying consumption due to the gas-fired generators being used to balance wind. While the forecasted consumption is globally balanced over longer time scales, the fluctuating consumption causes pressure fluctuations in the gas system to grow diffusively in time with a diffusion rate sensitive to the steady but spatially-inhomogeneous forecasted distribution of mass flow. To motivate our approach, we analyze the effect of fluctuating gas consumption on a model of the Transco gas pipeline that extends from the Gulf of Mexico to the Northeast of the United States.
\end{abstract}

\begin{IEEEkeywords}
Natural Gas Networks; Gas-Electric Coupling; Stochasticity; Reliability
\end{IEEEkeywords}

% For peer review papers, you can put extra information on the cover
% page as needed:
% \ifCLASSOPTIONpeerreview
% \begin{center} \bfseries EDICS Category: 3-BBND \end{center}
% \fi
%
% For peerreview papers, this IEEEtran command inserts a page break and
% creates the second title. It will be ignored for other modes.
\IEEEpeerreviewmaketitle

\section{Introduction}
\label{sec:intro}

A dominant new load on gas pipeline systems is natural gas-fired generators \cite{2010MITEI,2013MITEI}.   An example of this dramatic change is seen on the gas pipelines that supply the electrical grid controlled by the Independent System Operator of New England (ISO-NE) where natural gas-fired electrical generation increased from 5\% of total capacity to 50\% in a span of 20 years \cite{ISO-NE}.  A parallel development in many U.S. electrical grids is the expansion of intermittent renewable generation such as wind and photovoltaic (PV) generation---a trend that is expected to continue as utilities work to meet renewable portfolio standards \cite{portfolio_renewables,cost_renewables} that mandate a certain fraction of electrical generation be derived from renewable sources. In contrast to traditional coal, hydro or gas-fired generation, these intermittent renewable generators have limited controllability.  To maintain balance of generation and load, other grid resources must respond to counteract these new fluctuations.  Although many different types of advanced control of nontraditional resources are under consideration to provide balancing services, e.g. grid-scale battery storage and demand response, the control of fast-responding traditional generation (i.e. gas) is the current state-of-practice.

Gas pipelines have traditionally supplied Load Distribution Companies (LDC) that primarily serve space or water heating loads that evolve slowly throughout the day in a relatively well-known pattern that is predicted based on historical information and weather forecasts. Other traditional pipeline customers are industrial loads that change from day-to-day, but are very predictable over the span of twenty-four hours. The combination of expanded natural gas-fired generation and its use to balance intermittent renewable generation is creating loads on natural gas pipelines that are significantly different than historical behavior and will challenge the current pipeline operating paradigm that is used to control gas pressure.

The flow in natural gas pipeline is determined via bilateral transactions between buyers and sellers in a day-ahead market with market clearing and gas flow scheduling done in advance of the subsequent 24-hour period of gas delivery.  Scheduling consists of determining the locations and constant rates of gas injections. The initial market clearing assumes that gas consumptions are uniform over the subsequent 24-hour delivery period. Over the gas day, gas buyers improve their estimate of actual gas needs, and mid-course corrections are allowed through the transaction and scheduling of gas flows in two subsequent intra-day markets at 10 and 14 hours after the start of the 24-hour delivery period.

When serving traditional gas loads, the variability during the gas day is relatively small and slow and is well managed by linepack, i.e. compressed gas stored in the pipeline. The pressure in a gas transmission pipeline ranges between a maximum set by engineering limits and a minimum delivery pressure set by contracts.  A typical maximum pressure is around 800 psi, and flow of the gas causes the pressure to fall along the pipeline. As the minimum pressure ($\sim$ 500 psi) is approached, gas compressors installed along the pipeline are used to boost the pressure back near the maximum. Typical spacing between compressors is $\sim$ 50-100 km

The relatively high operating pressures enable large gas transfer rates, and the spread between maximum and minimum pressure allows the pipeline to operate with an imbalance of gas injections and consumptions for hours at a time.  An injection-consumption imbalance modifies the amount of gas stored in the pipeline via pressure changes, i.e. changes to the linepack.  Linepack is sufficient to buffer the imbalance when serving traditional gas loads. However, the hydraulic fracturing-driven expansion of natural gas-fired generation capacity \cite{10CWB} and its use to balance intermittent renewable fluctuations will result in larger and faster fluctuations in consumption (and possibly production) creating challenges to historical pipeline operations and reliability.

\begin{figure}
\centering
\includegraphics[width=0.45\textwidth]{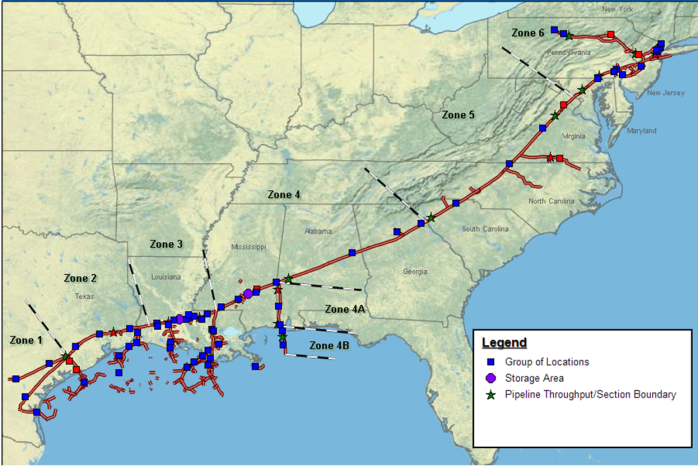}
\caption{Schematic representation of the Transco gas transmission network.}
\label{fig:Transco}
\end{figure}

The analysis in this manuscript is motivated by these new challenges.  Our approach is built on top of any solution of the steady gas flow problem that determines the spatial dependence of gas flow and pressure and the dispatch of gas compressors to maintain pressure.  For example, the steady flow solution can be found by solving an optimal gas flow (OGF) problem \cite{13MFBBCP} or using a model that approximates compressor dispatch decisions in current gas pipeline operations. Using these steady solutions, we build on the ideas of \cite{14CLB} and develop analysis tools to provide a probabilistic measure of the impact on pipeline reliability created by stochastic deviations of gas consumption from the forecasted values used in the steady solution (and scheduled during market clearing). These new tools are based on a linearization of the basic gas flow equations around the forecasted solution that retains the effect of stochasticity in consumption.  The effect of this stochasticity is assessed on a model of the Transco gas pipeline that extends from the Gulf of Mexico to the Northeast of the United States (See Fig.~\ref{fig:Transco} and  \cite{13MFBBCP}).

This manuscript builds on recent work \cite{13MFBBCP,14CLB} to develop a theoretic and computational approach to analyze the evolution of pressure in a gas system over time and space when the system is imbalanced. The three main contributions of this approach are:
\begin{itemize}
\item An analysis of spatiotemporal behavior of line-pack when the pipeline is subjected to stochastic gas consumptions. We observe that, even when fluctuations of the consumption and production are on average zero, the pressure fluctuations grow diffusively with time.
    %{\color{red} when the fluctuations of consumption are correlated in time}.
    We coin the term -- {\em diffusive jitter} of pressure fluctuations to describe this effect.

\item We show that the diffusive jitter of pressure is a nonlocal phenomenon where the pressure swings at one location depend on the behavior at all other locations.

\item We show that diffusive jitter is spatially inhomogeneous and dependent on the spatial distribution of the forecasted (stationary) solution.
\end{itemize}

The rest of the manuscript is organized as follows. Sections \ref{sec:TI} and \ref{subsec:DGF} provide a technical introduction to gas pipeline modeling. Section \ref{sec:OGF} provides a brief summary of approaches used to solve the steady gas flow problem. Section \ref{sec:pressure_fluct} describes a generalization of \cite{14CLB} that linearizes the gas flow equations around the steady solution that includes the effect of stochastic gas consumption.  The asymptotic solution of these linearized equations describes the diffusive jitter of the pressure fluctuations.
Section \ref{sec:exp} applies the theoretical results to a model of the Transco pipeline.  Section \ref{sec:con} summarizes our main results and offers a brief discussion of future work.

\section{Dynamic Gas Flow (DGF) Over a Single Pipe}
\label{sec:TI}
%\input{TI.tex}

%\subsection{Gas Flow Equations: Individual Pipe}
%\label{subsec:pipe}
Before analyzing a pipeline network, we introduce the gas flow equations and notations for a single pipe. Major transmission pipelines are typically 16-48 inches in diameter and operate at high pressures (e.g. $200$ to $1500$ psi) and high mass flows (millions of cubic feet of gas per day) \cite{Ref_Crane1982,Ref_Mokhatab2006}. Under these conditions, the pressure drop and energy loss due to shear is modeled by a nearly constant phenomenological friction factor $f$.  The resulting gas flow model is a nonlinear partial differential equation (PDE) with one spatial dimension $x$ (along the pipe axis) and one time dimension \cite{osiadacz1987simulation,87TT,05Sar}:
\begin{eqnarray}
&& \partial_t\rho+\partial_x (u\rho)=0,\label{density_eq}\\
&& \partial_t (\rho u)+\partial_x (\rho u^2)+\partial_x p=-\frac{\rho u |u|}{2d} f-\rho g \sin\alpha,\label{momenta_eq}\\
&& p=\rho Z R T.\label{thermodynamic_eq}
\label{state_eq}
\end{eqnarray}
Here, $u$, $p$, and $\rho$ are the spatially-dependent velocity, pressure, and density, respectively; $Z$ is the gas compressibility factor; $T$ is the temperature, $R$ is the gas constant, and $d$ is the diameter of the pipe.

Eqs.~(\ref{density_eq}, \ref{momenta_eq}, \ref{state_eq}) describe mass conservation, momentum balance and the ideal gas thermodynamic relation, respectively. The first term on the righthand side (rhs) of Eq.~(\ref{momenta_eq}) describes the friction losses in the pipe. The second term on the rhs of Eq.~(\ref{momenta_eq}) includes the gain or loss of momentum due to gravity $g$ when the pipe is tilted by angle $\alpha$.  The frictional losses typically dominate the gravity term, which is often dropped.  Because the flow velocities are usually small compared to the sound velocity, the gas inertia term $\partial_t(\rho u)$ and the advection term $\partial_x(\rho u^2)$ are typically small compared to the frictional losses and can also be dropped \cite{osiadacz1987simulation,87TT,05Sar}. For simplicity of presentation, we have also assumed that the temperature does not change significantly along the pipe.

Under these assumptions, Eqs.~(\ref{density_eq}, \ref{momenta_eq}, \ref{thermodynamic_eq}) are rewritten in terms of the pressure $p$ and the mass flux $\phi=u\rho$:
\begin{eqnarray}
&& c_s^{-2}\partial_t p + \partial_x\phi=0,\label{density_eq1.1}\\
&& \partial_x p+ \frac{\beta}{2d}\frac{\phi|\phi|}{p}=0,\label{momenta_eq1.1}
\end{eqnarray}
where $c_s\equiv \sqrt{ZRT}$  and $\beta\equiv f Z R T$ are considered constant. The solution of Eqs. (\ref{density_eq1.1}, \ref{momenta_eq1.1}) for $t\in[0,\tau]$ and $x\in[0,L]$ requires initial and boundary conditions,
\begin{eqnarray}
&& \forall x\in[0,L]:\quad \phi(0;x)=\phi_0(x),\label{phi-0}\\
&& \forall t:\quad \phi(t;0)=q^{(\mbox{\small in})}(t),\quad \phi(t;L)=q^{(\mbox{\small out})}(t),
\label{phi-in-out}
\end{eqnarray}
which are consistent, i.e. $\phi_0(0)=q^{(\mbox{\small in})}(0)$ and $\phi_0(L)=q^{(\mbox{\small out})}(0)$, in addition fixing the initial pressure at the beginning of the pipe, e.g. $p(0;0)=p_0$.

\section{Dynamic Gas Flow (DGF) Over a Network}
\label{subsec:DGF}

\begin{figure}
\centering
\includegraphics[width=0.48\textwidth]{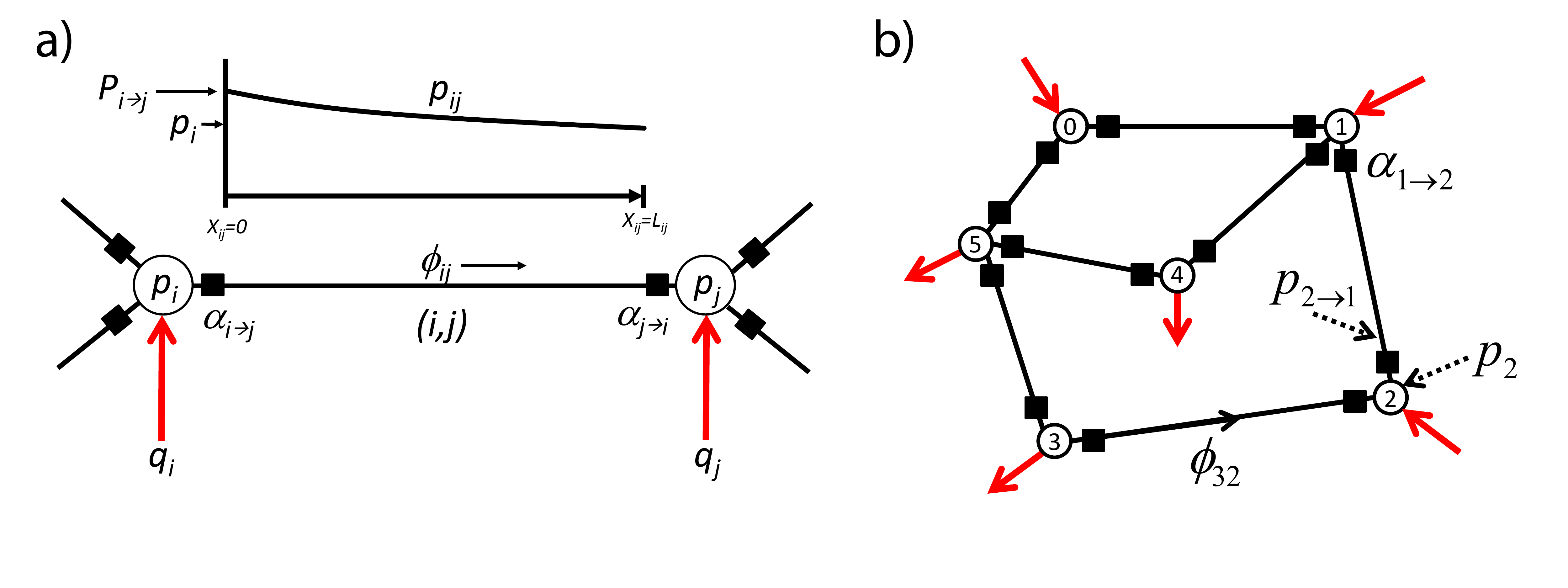}
\caption{Schematic illustration of a gas network and associated notation. a) A schematic illustration of a single edge $(i,j)$ of a network. Nodes at either end are indicated by open circles and labeled by their nodal pressure $p_i$ and $p_j$. Compressors are indicated with filled squares. Mass flow $\phi_{ij}$ is directed from $i$ to $j$ and nodal injections $q_i$ and $q_j$ contribute to this flow. Nodal pressure $p_i$ is modified by the compression ratio $\alpha_{i\to j}$ yielding $p_{ij}(x_{ij}=0)$.  The pressure falls along $\{i,j\}$ reaching $p_{ij}(x_{ij}=L_{ij})$.  If compressor $\alpha_{j\to i}$ is not present, then $p_{ij}(x_{ij}=L_{ij})=p_j$.  b) A schematic of many edges connected in a meshed network.  Nodes are indexed by $i=0,1,\cdots$, where node $0$ is typically reserved for the swing node -- the node where pressure is maintained constant throughout the dynamics. Compressors, injections and edge mass flows are the same as in a).
}
\label{fig:scheme}
\end{figure}

Next, we generalize the equations for a single pipe to a gas network. The network is modeled by a graph ${\cal G}=({\cal V},{\cal E})$ with vertices ${\cal V}$ and edges ${\cal E}$, where the edges are directed $(i,j)$ or undirected $\{i,j\}$ depending on the context. Each vertex $i\in{\cal V}$ represents a node with a gas mass injection or consumption rate $q_i$. Each edge $(i,j)\in{\cal E}$ is a single pipe with mass flow $\phi_{ij}$. The flow along each edge is described by a set of PDEs adapted from Eqs.~(\ref{density_eq1.1},\ref{momenta_eq1.1}):
\begin{eqnarray}
&&\hspace{-0.5cm}\forall t\in [0,\tau],\quad \forall \{i,j\}\in{\cal E},\quad \forall x\in[0;L_{ij}]:\nonumber\\ && c_s^{-2}\partial_t p_{ij}(t,x)+\partial_x \phi_{ij}(t,x)=0,\label{density_eq1}\\
&& \partial_x p_{ij}(t,x)+\frac{\beta}{2d} \frac{\phi_{ij}(t,x) |\phi_{ij}(t,x)|}{p_{ij}(t,x)}=0,\label{momenta_eq1}
\end{eqnarray}
where $p_{ij}(t,x)$ and $\phi_{ij}(t,x)$ are the pressure and mass flow, respectively, at time $t$ and position $x$ along edge $(i,j)$ of length $L_{ij}$. Here, $p_{ij}=p_{ji}$, $\phi_{ij}=-\phi_{ji}$, and $L_{ij}=L_{ji}$.  See Fig.~\ref{fig:scheme}a for a schematic description of the variables.

The flow of gas creates a pressure gradient, and compressor stations (potentially located at both ends of each edge $\{i,j\}$) are used to boost pressure.  $\alpha_{i\to j}$ denotes the compression ratio of the station adjacent to node $i$ that boosts pressure for flow toward node $j$ while $\alpha_{j\to i}$ denotes the compression ratio adjacent to node $j$ that boosts pressure for flow toward node $i$. We choose to place compressors at both ends of every edge for generality, which also simplifies the notations in the following discussion. In reality there is no more than one compressor on any particular edge of the graph and $\alpha=1$ when there is no compressor. Note that $\alpha_{i\to j}$ may be larger or smaller than 1, allowing the modeling of compression or decompression.  If only compression is allowed, then $\alpha_{i\to j}\geq 1$. The schematic in Fig.~\ref{fig:scheme} displays the spatial relationships between nodes, edges, and compression ratios. These are expressed mathematically as
\begin{eqnarray}
&&\hspace{-0.5cm}\forall t\in [0,\tau],\quad \forall (i,j)\in{\cal E}:\quad p_{ij}(t,0)=p_{i\to j}(t),\label{compressor}\\ && p_{ij}(t,L_{ij})=p_{j\to i}(t),\ \ p_{i\to j}=p_i \alpha_{i\to j},\ \ p_{j\to i}=p_j \alpha_{j\to i}, \nonumber
\end{eqnarray}
where $p_i$ and $p_{i\to j}$ are the pressures at node $i$ and pressure after compression ratio $\alpha_{i\to j}$ . If there is no compressor, then $\alpha_{i\to j}=1$. Under current operating practices, compression ratios do not change frequently \footnote{Compressors are automated to a degree in that they are run in modes with one of those modes considered in this manuscript being a constant compression ratio. Then pressure fluctuations at the inlet of a compressor are amplified at the outlet of that compressor.  Other modes of compressor operation may lead to different results for the fluctuation amplitude. There is relatively fast local control on the compressor to maintain this ratio. However, the set points for the ratio is updated very rarely. Adjusting this set point is the main effect the pipeline operator has on the system and this adjustment is done rarely throughout the operating day.}. Thus, we assume that $\alpha_{i\to j}$ does not depend on time.

Eqs.~(\ref{momenta_eq1},\ref{compressor}) are complemented with mass conservation at all nodes of the network:
\begin{equation}
\forall t\in [0,\tau],\quad \forall i\in{\cal V}: \sum_{j:(i,j)\in{\cal E}}\phi_{ij}(t,0)=q_i(t).
\label{flow_conserve}
\end{equation}
When the gas injections $q(t)=(q_i(t)|i\in{\cal V})$ are given for $t\in[0,\tau]$, nodal conditions (\ref{flow_conserve}) generalizes the single-pipe boundary conditions in (\ref{phi-in-out}) to a pipe network. Eqs.~(\ref{density_eq1}, \ref{momenta_eq1}, \ref{compressor}, \ref{flow_conserve}) constitute a complete set of equations describing the Dynamic Gas Flow (DGF) problem when they are supplemented with compression ratios, i.e. $\alpha=(\alpha_{i\to j}|(i,j)\in{\cal E})$, initial conditions on the flows
\begin{equation}
t=0,\ \forall \{i,j\}\in{\cal E},\ \forall x_{ij}\in[0,L_{ij}]:\quad \phi_{ij}(0;x_{ij})=\phi^{(in)}_{ij}(x_{ij}), \label{phi-initial-netw},
\end{equation}
and pressure at one slack node, $p_{i=0}(0)=p_0$.

%Ways of  setting the compressions are discussed in the next Section.

\section{Optimum Gas Flow Approaches}
\label{sec:OGF}

Later in the manuscript, we analyze the DGF problem by linearization of the fluctuations about a stationary solution.  Here, we summarize two approaches to finding this stationary solution.  We first solve a stationary version of the DGF problem, i.e, the Gas Flow (GF) problem, where the steady state pressure and flows are expressed in terms of compression ratios.  The time-independent compression ratios are then determined via solution of the OGF problem.

\subsection{Stationary Gas Flow}
\label{sec:steady}

In the GF problem, all input parameters (consumptions/injections, compression ratios and the pressure at the slack bus) are constant in time. The total injection and consumption is balanced
\begin{equation}
\sum_{i\in{\cal V}} q_i^{(\mbox{st})}=0. \label{ballance}
\end{equation}

The steady solution of Eq.~(\ref{density_eq1}) is uniform mass flow along each pipe in the network, $\forall \{i,j\}:\quad \phi_{i\to j}=\mbox{const}$. Substituting this result into Eq.~(\ref{momenta_eq1}) and integrating over space yields the algebraic relationship between pressure at position $x\in[0;L_{ij}]$, compression, and (constant) flow through the pipe
\begin{eqnarray}
&& \hspace{-0.5cm}\forall (i,j)\in{\cal E}:\quad p_{i\to j}^{(\mbox{st})}=p_i^{(\mbox{st})} \alpha_{i\to j}; \nonumber\\ &&
(p_{ij}^{(\mbox{st})}(x))^2=(p_{i\to j}^{(\mbox{st})})^2-\frac{\beta x}{d} \phi_{ij}^{(\mbox{st})}|\phi_{ij}^{(\mbox{st})}|.
\label{p-phi-steady}
\end{eqnarray}
The GF problem has a unique solution provided the compression ratios are known, and the GF solution in (\ref{p-phi-steady}) is the basis for many approaches to solving the OGF problem.

\subsection{Optimum Gas Flow}

The solution to the GF problem leaves the time-independent compression ratios $\alpha$ unknown.  These are chosen by the pipeline operators based on a combination of economic and operational factors. Here, we describe two approaches for selecting the $\alpha$.  The first is a greedy algorithm that approximates the current pipeline operations in the US.  The guiding principle is that compressors are activated when the pressure prior to the next compressor drops below the acceptable lower bound.  When activated, a compressor is set to its maximum compression ratio $\overline{\alpha}$. This algorithm is described in detail in \cite{13MFBBCP}. The second approach is based on solutions to the optimal gas flow (OGF) problem \cite{68WL,00WRBS,10Bor,13MFBBCP}.  Here, we summarize a Geometric Programming (GP) approach to solving the OGF problem that minimizes the total compressor power to move the gas.
%This algorithm is also described in detail in \cite{13MFBBCP}. {\color{blue} Just making sure he describes these binary variables in his thesis}

The total power used in pipeline gas compression (assuming that the gas is ideal and compression is isentropic) is
\begin{eqnarray}
\sum_{(i, j) \in \cal{E}} \frac{ c_{i \to j} \phi_{ij}^{(st)}} {\eta_{i \to j}} { \left( \max\{\alpha_{i \to j}^m,1\}-1\right)},%  \left(\alpha_{i \to j}^m-1\right),
\label{cost}
\end{eqnarray}
where $c_{i \to j}$ is a constant that depends on the compressor, $m = (\gamma - 1)/ \gamma$ where $\gamma$ is the gas heat capacity ratio, and $\eta_{i \to j}$ is the efficiency factor of the compressor. It is important to note that fluctuations caused by compressor consumption are negligible when compared to gas loads. The term $\phi_{ij}^{(st)}$ denotes the {directional mass} flow for edge $i,j$, when the edge is oriented from $i$ to $j$. The OGF formulation assumes that
the flow through the compressor is from $i$ to $j$, i.e., $\phi_{ij}^{(st)} > 0$, thus the direction of flow must be selected before hand. For tree networks, the magnitude and direction of the flows are computed exactly {\it apriori} and do not depend on the choice of compression ratios. For an edge $i,j$, let $\mathcal{G}_i$ and
$\mathcal{G}_j$ be the two disjoint graphs obtained by removing $(i,j)$. The flows $\phi_{ij}^{(st)}$ are computed as %The flow direction $\phi_{ij}^{(st)}$ is computed as
\begin{align} \label{flowcomputation}
	\phi_{ij}^{(st)}=\sum_{i\in\mathcal{G}_i} q_i^{(st)}=-\sum_{i\in\mathcal{G}_j} q_i^{(st)}.
\end{align}
In networks with loops, flow direction is chosen using heuristics or through the introduction of binary variables \cite{10Bor}.

Using the cost function in Eq.~(\ref{cost}), the OGF problem is formulated as
\begin{eqnarray}
\min_{\alpha, p} & &  \sum_{(i,j)\in{\cal E}} \frac{ c_{i \to j} \phi_{ij}}{\eta_{i \to j}}{ \left( \max\{\alpha_{i \to j}^m,1\}-1\right)} \label{obj} \\ %\left( \alpha_{i \to j}^m-1\right)^+  \label{obj}
\mbox{s.t.} & & \forall (i,j)\in{\cal E}:\quad \alpha_{i \to j}^2 =
\frac{p_j^2 + \frac{\beta L_{ij}}{d_{ij}}  \phi_{ij}^2}{p_i^2 }, \label{compression_ratio} \\
&& \forall i\in{\cal V}:\quad 0\leq \underline{p}_i\leq p_i\leq \overline{p}_i,
\label{p-box}\\
&& \forall (i,j)\in{\cal E}:  {\underline{\alpha}_{i \to j}} \leq  \alpha_{i \to j}\leq \overline{\alpha}_{i \to j}, \label{alpha-box}
\end{eqnarray}
where Eq.~(\ref{compression_ratio}) is obtained from Eq.~(\ref{p-phi-steady}).
The upper bound in Eq.~(\ref{p-box}) represents engineering limits on pipes and the lower bound represents contractual obligations. The upper and lower bounds in Eq.~(\ref{alpha-box}) refer to maximum allowed compression and decompression at each compressor.  If decompression is not allowed, $\underline{\alpha}_{i \to j}=$1.

There are a variety of methods for solving the OGF over trees, and in this paper, we use the geometric programming (GP) approach described in  \cite{13MFBBCP}. The GP approach relaxes the lower bound  $\underline{\alpha}_{i \to j}$ in Eq.~(\ref{alpha-box}), i.e. $\underline{\alpha}_{i \to j}=$0. Under this relaxation, the OGF is transformed into a GP of the form
 \begin{align}
 	\min_{\hat{t}, \hat{\beta}} \quad & \log \left( \sum_{(i,j) \in \mathcal{E}} d_{ij} e^{m \hat{t}_{ij}}  \right), \quad \forall i \in \mathcal{V} \label{geombegin} \\
	\mbox{s.t. } \quad & 2	\log(\underline{p}_i) \leq \hat{\beta}_i \leq 2 \log(\bar{p}_i) \\
	& 0 \leq  \hat{t}_{ij} \leq \log(\bar{\alpha}_{ij}), \\
	& \log \left( e^{\hat{\beta}_j - \hat{\beta}_i-\hat{t}_{ij}} + \delta_{ij}^1 e^{-\hat{\beta}_i - \hat{t}_{ij}}  \right) \leq 0,  \label{geomend} \\
	& \qquad \mbox{   \hspace{1.5in}} \forall (i,j) \in \mathcal{E}. \nonumber
  \end{align}
  The transformed variables are related to the original ones via the following equations
  \begin{align}
  	p_i^2 = e^{\beta_i}, \quad \delta_{ij}^1 = \frac{\beta L_{ij}}{d_{ij}}  \phi_{ij}^2.
  \end{align}
The OGF in Eqs.~(\ref{geombegin}-\ref{geomend}) is solved using convex optimization. When decompression is not allowed, $\underline{\alpha}_{i \to j}$=1 in Eq.~(\ref{alpha-box}), we use a signomial programming (SP) method, which is a heuristic version of GP based on solving a sequence of convex programs \cite{13MFBBCP}.  Here, we use SP to solve the OGF.

\section{Diffusive Jitter of Pressure Fluctuations}
\label{sec:pressure_fluct}

The main contribution of this manuscript builds on the solution of the OGF by introducing a model of stochastic gas consumption and the analysis of its effects on pressure fluctuations--{\it diffusive jitter}.  Our approach linearizes the DGF equations (Eqs.~\ref{density_eq1},\ref{momenta_eq1}) around a solution to the GF problem (augmented with compression ratios from the OGF or the greedy compression scenario).

The linearized model captures the relationship between the fluctuating consumption and the fluctuating pressure. Asymptotically, the accumulated changes in pressure provide an indication of how fast the pressure will drift (the jitter) and exceed an operating limit in the absence of operator intervention. As the DGF solution drifts further from the original GF solution, the quality of the linearization degrades.  However, we expect that the linearized solution to the DGF remains a strong relative indicator of how quickly a system will experience problems due to stochastic consumption.

%The GF and OGF problems of the previous sections apply to gas pipelines with well behaved loads.
Formally, the stochastic consumption is defined by $q(t)=q^{(\mbox{st})}+\xi(t)$ where the components of $\xi(t)=(\xi_i(t)|i\in{\cal V})$ are time varying but relatively small in comparison to $q^{(\mbox{st})}$. We assume a linearized solution of the DGF problem of the form $p(t)=p^{(\mbox{st})}+\delta p(t)$ and $\phi(t)=\phi^{(\mbox{st})}+\delta \phi(t)$,  where the respective corrections are small, i.e. $|\delta p(t)|\ll p^{(\mbox{st})}$ and $|\delta \phi(t)|\ll \phi^{(\mbox{st})}$. The linearized versions of Eqs.~(\ref{density_eq1}, \ref{momenta_eq1}, \ref{compressor}, \ref{flow_conserve}) are
\begin{eqnarray}
&&\hspace{-0.5cm}\forall t\in [0,\tau],\quad \forall \{i,j\}\in{\cal E},\quad \forall x\in[0;L_{ij}]:\nonumber\\
&&c_s^{-2}\partial_t \delta p_{ij}+\partial_x \delta \phi_{ij}=0,\label{density_delta}\\
&& \partial_x \delta p_{ij}+\frac{\beta}{2d} \Biggl(\frac{\delta\phi_{ij} |\phi_{ij}^{(\mbox{st})}|}{p_{ij}^{(\mbox{st})}}+\nonumber\\ &&
\frac{\phi_{ij}^{(\mbox{st})} |\delta\phi_{ij}|}{p_{ij}^{(\mbox{st})}}-\frac{\delta p_{ij}\phi_{ij}^{(\mbox{st})} |\phi_{ij}^{(\mbox{st})}|}{(p_{ij}^{(\mbox{st})})^2}\Biggr)=0,\label{momenta_delta}\\
&&\hspace{-0.5cm}\forall t\in [0,\tau],\quad \forall (i,j)\in{\cal E}:\nonumber\\
&& \delta p_{i\to j}=\delta p_i \alpha_{i\to j} , \label{compressor_delta}\\
&& \delta p_{ij}(t,0)=\delta p_{i\to j}(t),\quad \delta p_{ij}(t,L_{ij})= \delta p_{j\to i}(t),\label{conditions_delta_p}\\
&&\hspace{-0.5cm}\forall t\in [0,\tau],\quad \forall i\in{\cal V}:\quad
\sum_{j:(i,j)\in{\cal E}}\delta \phi_{ij}(t,0)=\xi_i(t).
\label{conditions_delta_phi}
\end{eqnarray}
We seek asymptotic solutions to the PDE of Eqs.~(\ref{density_delta}, \ref{momenta_delta}, \ref{compressor_delta},\ref{conditions_delta_p},\ref{conditions_delta_phi}), where asymptotic implies finding solutions for time $\tau$ longer than the correlation time of the fluctuation consumption $\xi$. In addition, we seek solutions of Eqs.~(\ref{density_delta}, \ref{momenta_delta},\ref{compressor_delta},\ref{conditions_delta_p},\ref{conditions_delta_phi}) that connect the nodal quantities by algebraic relationships thereby eliminating the complexity of the orginal PDE.

The solution approach is an extension of the work in \cite{14CLB}.  Following \cite{14CLB}, we solve Eqs.~(\ref{density_delta}, \ref{momenta_delta}) for each pipe using a proposed solution of the form
\begin{equation}
\delta p_{ij}=a_{ij}(t)Z_{ij}(x)+b_{ij}(t,x),\label{delta-p}
\end{equation}
where the $a_{ij}(t)$ depend on time. In \cite{14CLB}, it was argued that the $a_{ij}(t)Z_{ij}(x)$ term represents the asymptotic contribution to the  gas pressure fluctuations that grows in time.  In contrast, $b_{ij}(t,x)$ represents smaller contributions to the pressure fluctuations that do not grow in time. Here, we focus on the contribution from the $a_{ij}(t)Z_{ij}(x)$ term which is asymptotically dominant at long times.\footnote{Bounds for the second term are derived and solved using inhomogeneous linear equations for $b_{ij}$.}

Substitution of proposed solution (\ref{delta-p}) into Eqs.~(\ref{density_delta}, \ref{momenta_delta}) yields an equation for $Z_{ij}$, i.e.
\begin{equation}
\partial_x Z_{ij}-\frac{\beta}{2d}\frac{\phi_{ij}^{(\mbox{st})}|\phi_{ij}^{(\mbox{st})}|}{(p_{ij}^{(\mbox{st})})^2}Z_{ij}=0,
\label{Z}
\end{equation}
where $Z_{ij}(x)$ counts $x$ from node $i$. The integration of Eq.~(\ref{Z}) over the spatial dependence of the stationary profile (\ref{p-phi-steady}),  yields
\begin{eqnarray}
Z_{ij}(x)=\frac{p_{i\to j}^{(\mbox{st})}+p_{j\to i}^{(\mbox{st})}}{2p_{ij}^{(\mbox{st})}(x)},
\label{Z2}
\end{eqnarray}
where the normalization constant is chosen to guarantee, $\int_0^L Z_{ij}(x) dx/L=1$.

We solve for the time-dependent factor $a_{ij}(t)$ by substituting $\delta p_{ij}$ $\sim$ $a_{ij}(t)Z_{ij}(x)$ into Eq.~(\ref{density_delta}) and integrating the result over the entire spatial extent of the pipe $\{i,j\}$ yielding
\begin{eqnarray}
&& a_{ij}(t)=c_s^2\int_0^tdt'\left(\delta\phi_{ij}(t',0)-\delta\phi_{ij}(t',L)\right).
\label{a-via-delta-phi}
\end{eqnarray}

In the asymptotic limit where $\delta p_{ij}$ $\sim$ $a_{ij}(t)Z_{ij}(x)$ for every pipe (graph edge), Eqs.~(\ref{conditions_delta_p}) can only be satisfied if the $a_{ij}(t)$ have the same functional dependence on time, i.e.,
\begin{equation}
\forall \{i,j\}\in{\cal E}:\quad a_{ij}(t)=a(t)c_{ij},\label{a}
\end{equation}
where $c_{ij}=c_{ji}$ is an edge specific constant.

To compute the global time-dependent factor $a(t)$ we sum the mass conservation equation over all the nodes of the graph
\begin{equation}
\sum_{i\in{\cal V}}\xi_i=\sum_{\{i,j\}\in{\cal E}}\left(\delta\phi_{ij}(t,0)-\delta\phi_{ij}(t,L_{ij})\right),
\label{sum_delta_phi}
\end{equation}
integrate over time and define
\begin{equation}
\Xi(t)\doteq\int_0^tdt' \sum_{i\in{\cal V}}\xi_i(t'),\label{Xi}
\end{equation}
and finally sum Eq.~(\ref{a}) overall edges:
\begin{equation}
a(t)=\frac{c_s^2 \Xi(t)}{\sum_{\{i,j\}\in{\cal E}}c_{ij}}.
\label{aa}
\end{equation}
Therefore, $\forall t,\ \ \forall \{i,j\}\in{\cal E},\ \ x\in[0,L_{ij}]:$
\begin{equation}
\delta p_{ij}(t,x)\approx \frac{c_s^2\Xi(t)}{\sum_{\{i,j\}\in{\cal E}}c_{ij}} c_{ij}Z_{ij}(x).\label{delta_p_final}
\end{equation}
The unknown edge constants $c_{ij}$ are derived by substituting Eqs.~(\ref{delta_p_final}) into Eqs.~(\ref{compressor_delta}, \ref{a}) yielding
\begin{equation}
\forall i,\ \ \forall j,k\ \ \mbox{s.t.  }(i,j),(i,k)\in{\cal E}:\ \ \frac{c_{ij} Z_{ij}(0)}{\alpha_{i\to j}}=\frac{c_{ik}Z_{ik}(0)}{\alpha_{i\to k}}.
\label{c-relations}
\end{equation}
Eqs.~(\ref{delta_p_final}, \ref{c-relations}, \ref{Z2}) express the complete asymptotic (zero mode) solution of the DGF problem.

Finally, we make several observations to connect the solution for the pressure fluctuations in Eqs.~(\ref{delta_p_final}, \ref{c-relations}, \ref{Z2}) to a probability distribution over the pressure fluctuations.  First, the random gas load fluctuations $\xi_i(t)$ are zero-mean, temporarily homogeneous, and  relatively short correlated in both time (the correlation time is less than $\tau$) and space (the correlation length is less than the spatial extent of the network). Second, the fluctuations of $\delta p_{ij}$ in Eq.~(\ref{delta_p_final}) are given by a time-integral and spatial-sum of the fluctuations.  According to the Large Deviation theory, these observations imply that {\em the pressure fluctuations form a Gaussian random process which jitters diffusively in time}. Specifically, the Probability Distribution Function (PDF) of $\delta p_{ij}(t,x)$ is
\begin{eqnarray}
&& \hspace{-1cm}{\cal P}(\delta p_{ij}(t,x)=\delta)\to \left(2\pi t D_{ij}(x)\right)^{-1/2}
\exp\left(-\frac{\delta^2}{2 t D_{ij}(x)}\right),
\label{PDF}\\
&& \hspace{-1cm} D_{ij}=\left(\frac{c_s^2 c_{ij}Z_{ij}(x)}{\sum_{\{k,l\}\in{\cal E}}c_{kl}}\right)^2
\Biggl\langle \left(\sum_{n\in{\cal V}}\xi_n(t')\right)^2\Biggr\rangle
,\label{xi-xi}
\end{eqnarray}
where the correlation function on the right-hand-side does not depend on $t'$  due to assumption of the statistical homogeneity of $\xi$.

\section{Numerical Experiments}
\label{sec:exp}

Inspection of Eq.~(\ref{xi-xi}) shows that the variance of the pressure fluctuations as a function of position in the network is related to the coefficients $D_{ij}(x)$, referred to collectively as $D$. Higher values of $D$ correspond to larger pressure fluctuations and higher likelihood of the pressure violating an engineering or contractual limit.  By analogy with related physical processes, the coefficients $D$ are similar to a diffusion coefficient, and we refer them this way in the remainder of the manuscript.  The origins of $D$ are primarily twofold.  Once the gas consumptions and injections are fixed, the spatial dependence of $D$ arises from the particular stationary solution of pressures, flows, and compression ratios through the $Z_{ij}(x)$. The magnitude of $D$ is also related to the average global strength of the consumption fluctuations $\langle \left(\sum_{n\in{\cal V}}\xi_n(t')\right)^2\rangle$.

We apply the results described above to the Transco pipeline shown schematically in Fig.~\ref{fig:Transco}.  We use data for the total consumption at each node over a 24-hour period from December 29, 2012 to fix the forecasted consumption for the stationary GF solution. %{\color{red} Michael, confirm that you used average over 24-hours, or if this was hourly data)}.
These data represent relatively stressed operations for the Transco pipeline.  The Transco pipeline has a small number of loops, which we partition to create a tree topology \cite{13MFBBCP} that is very nearly linear but with a few small branches.  We resolve these branches in the solution of the GF (or OGF) problem, however, when analyzing the pressure fluctuations, we aggregate these short branches to nodal consumptions and only analyze the fluctuations as a function of distance along the mainline.

The Transco operational data does not include information on the deviations of the gas flows from their average or scheduled values. %{\color{red} Michael, please verify}.
Instead, we estimate the global mean-square consumption fluctuations as
\begin{equation}\label{eq:fluc_estimate}
\left\langle \left(\sum_{n\in{\cal V}}\xi_n(t')\right)^2\right\rangle \approx
\left(\frac{\phi_0}{3}\right)^2*N.
\end{equation}
Here, $\phi_0 \approx 20 \mbox{ kg/s}$ is a typical average consumption for a node in the Transco pipeline, and $N \approx 70$ is the number of consumption nodes, e.g. city-gates or power plants. This estimate of the gas consumption fluctuations assumes that the fluctuations at neighboring nodes are uncorrelated. If these neighboring nodes are gas-fired turbine generators that are both being used to balance renewable fluctuations, the assumption of independence may lead to an underestimation in Eq.~(\ref{eq:fluc_estimate}).

For presentation purposes, it is convenient to find a suitable normalization for $D$.  Motivated by Eqs.(\ref{PDF},\ref{eq:fluc_estimate}), we normalize $D$ by
\begin{align*}
D_0 \approx \left(\frac{p_0}{3}\right)^2/t_0
\end{align*}
where $p_0=800 \mbox{ psi}\approx 5.5*10^6 \mbox{ Pa}$ is the upper
bound on allowed pressure in the pipes and $t_0=15\mbox{ min}\approx 10^3
s$ is a representative time period where we expect the developed theory to
work well.

\begin{figure}
\centering
\includegraphics[width=0.45\textwidth]{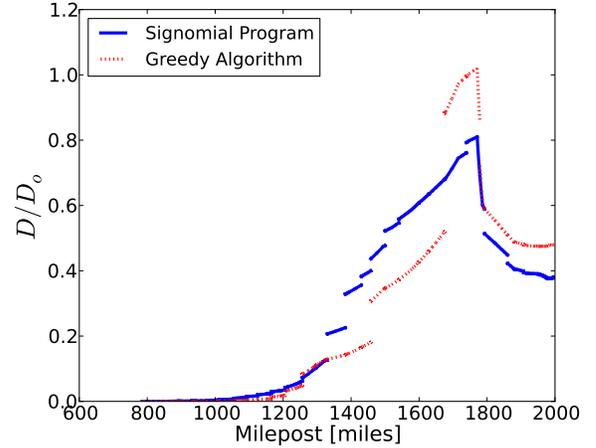}
\caption{Diffusion coefficient as a function of distance along the Transco
mainline with stationary solutions given by OGF and the
greedy algorithm.  Both show a peak at milepost $1771$, but the magnitude
of this peak is much higher for the greedy algorithm than the OGF, indicating larger pressure fluctuations for the greedy algorithm.
}
\label{BBoth}
\end{figure}

We consider the base case of December 29th, 2012 and several modifications of this base case to investigate the effects of changing operations.   Fig.~\ref{BBoth} displays $D$ as a function of location along the mainline for two different stationary solutions for the base case---the OGF solution described in Section~\ref{sec:OGF} and the greedy algorithm from \cite{13MFBBCP}. For a characteristic time of $15\mbox{ min} \approx 10^3s$, \; $D/D_o = 1$ corresponds to a variance in pressure fluctuations of $(266\mbox{ psi})^2$. Pressures in the Transco Pipeline range between $500\mbox{ psi}$ and $800\mbox{ psi}$, so the pressure fluctuation standard deviation is $33-53\%$ of the pressures in the pipeline for $D/D_o = 1$. The same characteristic time and $D/D_o = 0.1$ yields a variance of $(84\mbox{ psi})^2$ which gives pressure fluctuation standard deviations of $10-16\%$ of pipeline pressures. Since the pressure variance grows linearly in time, often over several $15\mbox{ min}$ intervals, these fluctuations can quickly grow to exceed pressure bounds without proper intervention. As the plots show, most pressure fluctuations are above $D/D_o = 0.1$ throughout the pipeline, and therefore the fluctuations are of concern in any regions of pipeline where the pressure is near its upper or lower bound. The two solutions display similarities. Both show a build up of $D$ from milepost 800 nearer to the Gulf of Mexico, a peak at milepost 1771 near New York and New Jersey, and a decay to a smaller value at milepost 2000 near the injection point for the Marcellus Shale in Pennsylvania.

The spatial variation of $D$ is due to $Z_{ij}(x)$, and the general shape of $D$ can be understood by revisiting Eq.~\ref{Z}.  The form of this equation suggest exponential growth or decay of $Z_{ij}(x)$ depending on the orientation of $\phi_{ij}$. The flow from the Gulf to the New York/New Jersey area is unidirectional creating the growth of $D$ observed in Fig.~\ref{BBoth}.  However, the large loads in the New York/New Jersey area combined with the offsetting injections from from the Marcellus Shale creates a flow reversal and an exponential decay of $Z_{ij}$ (and therefore of $D$).  The peak in $D$ is connected to the point of flow reversal.  The solution in Fig.~\ref{BBoth} displays more structure than simple exponential growth and decay for several reasons.  First, the mass flow rates $\phi_{ij}$ depend on location.  However, perhaps more important are the discontinuities in $D$.  These occur at compressor stations and are due to the discontinuities in $p^{\mbox{(st)}}$ at these locations.

The global behavior of the OGF solution and the greedy algorithm  Fig.~\ref{BBoth} is similar because the differences in the compression ratios in the stationary solution does not affect the mass flow rates.  However, it does affect the spatial dependence of pressure which can lead to the substantial local differences observed in Fig.~\ref{BBoth}.
%We compare the profile of $D$ for the stationary solutions given by the
%OGF and the greedy algorithm, as shown in Fig.~\ref{BBoth}.
%The peak $D$ is the same for both algorithms, since this is point of flow reversal in both cases.
%In addition, there are no other local maxima and the magnitude of $D$
%is much lower near the Gulf than in Pennsylvania for both.
For example, between mileposts 1400 and 1700, $D$ is much lower for the greedy algorithm compared to the OGF. However, the deployment of a compressor near milepost 1700 in the greedy algorithm leads to a large jump in $D$, a larger peak in pressure fluctuations, and a greater chance for violation of an engineering or contractual pressure limit.  For this one example, this difference would seem to suggest that the OGF solution is less susceptible to pressure fluctuations.  However, we note that the expected magnitude of the pressure fluctuations is not taken into account in either the greedy algorithm or the OGF.

What these results do suggest is that the deployment of compressors in the stationary solution can have a significant impact on the expected pressure fluctuations, and that it is possible to formulate a a compressor dispatch optimization that balances the risk of such fluctuations against other desired operational properties, e.g. cost.  The simple algebraic form of the probability of such large fluctuations in Eqs.~(\ref{PDF},\ref{xi-xi}) are convenient for incorporation into such formulations.

%The plot of $D$ for the greedy algorithm has far fewer discontinuities than
%for the OGF since it only operates compressors when the
%pressure bounds would otherwise be violated and it operates each compressor
%at its maximum allowed compression ratio, leading to fewer compressors in
%operation than for the OGF

% Way too strong to say this - rbent
%({\color{cyan} requires further clarification or
%should be removed?}).
%We conclude that reducing pressure fluctuations requires operating
%more compressors than are used in the greedy algorithm, and comes with a
%reduction in cost of compression ({\color{cyan} too much?}).

\begin{figure}
\centering
\includegraphics[width=0.45\textwidth]{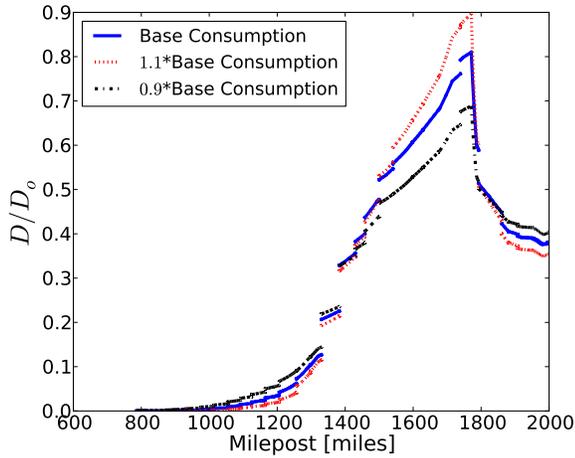}
\caption{Diffusion coefficient as a function of distance along the Transco
mainline with stationary solutions given by the OGF with
global consumption and injection scaled by a uniform factor.
All show a peak at milepost $1771$, but higher scaling factors have higher
magnitudes at their peaks, indicating larger pressure fluctuations for larger
system loads.
}
\label{SSig}
\end{figure}

To determine the effect of overall consumption and injection on $D$,
we uniformly scaled the base case consumptions and injections by a constant factor---a scaling that preserves the balance of consumptions and injections required for the existence of a stationary solution. Figure~\ref{SSig} displays the results for $D$ computed using the OGF stationary solutions.  The stationary solutions show small local differences in $D$ caused by the deployment of compression. However, the major impact stems from the increase (or decrease) in mass flows.  The uniform scaling does not affect the location of the flow reversal, so the peak in $D$ appears at the same place.  However, the larger (smaller) flows lead to faster (slower) growth rates for $Z_{ij}$ (see Eq.~\ref{Z}) and an overall higher (lower) peak in $D$.

\begin{figure}
\centering
\includegraphics[width=0.45\textwidth]{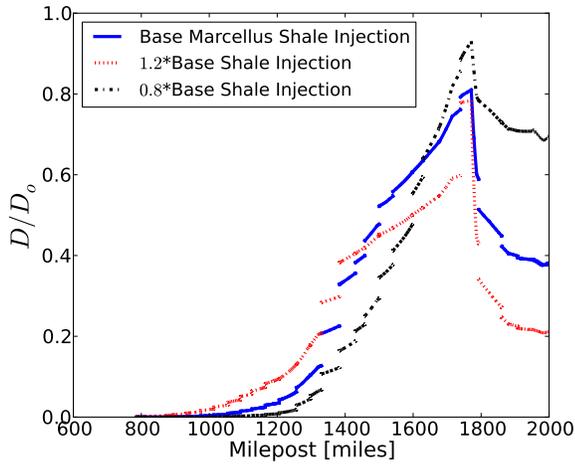}
\caption{Diffusion coefficient as a function of distance along the Transco
mainline with stationary solutions given by the OGF with
Marcellus Shale injections scaled by a factor and the corresponding amount
of injections removed from the Gulf.  All show a peak at milepost $1771$, but
higher scaling factors have slightly lower magnitude at the peak, indicating
smaller pressure fluctations.
Higher scaling factors also have much lower magnitudes in the Marcellus
Shale, indicating smaller pressure fluctuations when injections are shifted
from the Gulf to the Marcellus Shale.
}
\label{SuSig}
\end{figure}

In recent years the Marcellus Shale has become a large supplier of
gas, and its injection capability is expected to increase \cite{NERC2013-gas-grid}.
To model the effect of this expansion we scaled all injections from the
Marcellus Shale by a constant factor and removed a corresponding amount of
gas from the injections at the Gulf to preserve the global balance
of consumption and injection.  Fig.~\ref{SuSig} displays the results for $D$ along the Transco pipeline for the OGF solution. Although the injection from the Marcellus is increased, the major gas load centers in New York and New Jersey keep the flow reversal point, and therefore the peak in $D$, pinned at more or less the same location. Larger Marcellus injections show slightly lower peak amplitudes of $D$ indicating that, as gas injections are shifted from the Gulf to the
Marcellus Shale, the reliability of pipeline operations is improved. We conjecture that this change is due to moving the source of gas injections closer to the major load centers.  However, we again note that the OGF methods used to find the stationary solution do not account for the expected pressure fluctuations and their inclusion will likely lead to a modification of these results.

\begin{figure}
\centering
\includegraphics[width=0.45\textwidth]{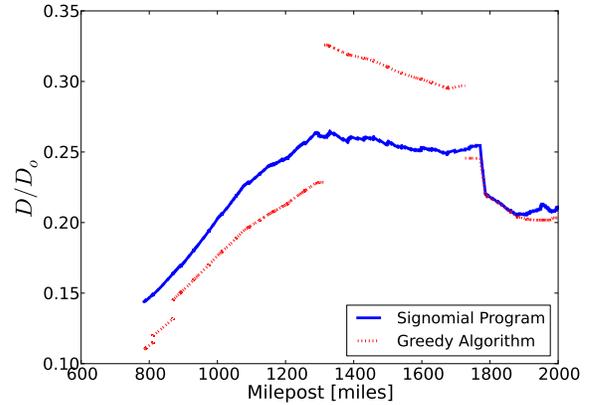}
\caption{Diffusion coefficient as a function of distance along the Transco
mainline with load redistributed from the large load in New York to
the Gulf and Marcellus Shale, leaving the large load in New Jersey unaltered.
This causes the appearance of a new global maximum at milepost $1319$
which is the location of a large load in North Carolina.
Since the New Jersey load was not redistributed, a local maximum remains
at milepost $1771$.
}
\label{shift1}
\end{figure}

Although scaling overall consumption and Marcellus supply are directly relevant to operators and planners, they do not exhibit a shift in the location of the peak in $D$ or the appearance of multiple local maxima. To study these possibilities, we next imposed some less realistic changes.
In particular, the loads in New York are shifted to points closer to the Gulf and Marcellus Shale, but the New Jersey loads were left unaffected.  Fig.~\ref{shift1} displays the impact of this shift on $D$ computed using stationary solutions from the OGF and the greedy algorithm.
Since the large load in New Jersey remains, milepost $1771$ is still a position
of flow reversal and a, now minor, maximum of $D$. However, the redistribution of load leads to a new global maximum near milepost $1319$---a large load in North Carolina.  Although the location of maximum fluctuations has been relocated, the maximum of $D$ is much reduced by moving the loads closer to the gas injections.

\begin{figure}
\centering
\includegraphics[width=0.45\textwidth]{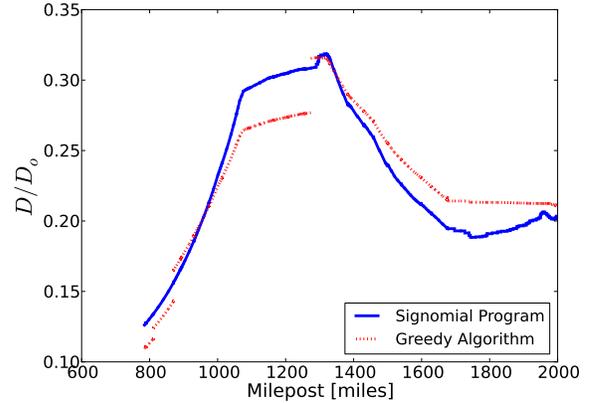}
\caption{Diffusion coefficient as a function of distance along the Transco
mainline with load redistributed from the large loads in New York and New
Jersey closer to the Gulf and Marcellus Shale.
The global maximum at milepost $1319$ remains, but the local maximum
at milepost $1771$ dissappears since the large load has been removed from
this area.
}
\label{shift2}
\end{figure}

%{\color{red} Might consider deleting the next paragraph and Fig. 7.  They don't add that much to the discussion...... }
The previous example showed the appearance of a new global maximum, as well as several small local maxima, but the original local maximum remained.  To remove it, we shifted New Jersey's load to points closer to the Gulf and the Marcellus Shale, as shown in Fig.~\ref{shift2}.
%in the code this is really a redistribution of injection from the Gulf
%and Marcellus Shale into the New Jersey branch
This successfully removes the local maximum at milepost $1771$ while leaving the global maximum at milepost $1319$. In this case the jitter (diffusion coefficient) of the greedy algorithm and OGF
are comparable, with OGF jitter greater before milepost $1319$ and greedy greater after milepost $1319$.

%\begin{figure}
%\centering
%\includegraphics[width=0.45\textwidth]{Figures/Results/base_signomial.png}
%\caption{Diffusion coefficient as a function of distance along the Transco
%mainline with stationary solution given by OGF.
%The peak occurs at milepost $1771$, which is where the gas flowing from the
%Gulf joins the Marcellus Shale gas and together flow down two small branches
%to service large loads in New Jersey and New York.
%}
%\label{BSig}
%\end{figure}

%% \begin{figure}
%% \centering
%% \includegraphics[width=0.45\textwidth]{Figures/Results/base_greedy.png}
%% \title{Greedy Algorithm for the Transco Pipeline}
%% \caption{To be written... }
%% \label{BGr}
%% \end{figure}

%% \begin{figure}
%% \centering
%% \includegraphics[width=0.45\textwidth]{Figures/Results/scale_greedy.png}
%% \title{Diffusion Versus Scaling of Global Consumption for Greedy Algorithm}
%% \caption{To be written... }
%% \label{SGr}
%% \end{figure}

%% \begin{figure}
%% \centering
%% \includegraphics[width=0.45\textwidth]{Figures/Results/supply_greedy.png}
%% \title{Diffusion Versus Scaling of Marcellus Shale Injection for Greedy
%% Algorithm}
%% \caption{To be written... }
%% \label{SuGr}
%% \end{figure}

\section{Conclusions and Future Work}
\label{sec:con}

We have focused the analysis on the coupling between natural gas networks and electric networks at the time scale of intra-day natural gas markets.  The coupling at this time scale is expected to become tighter because of several factors: the retirement of coal and fuel oil-fired generation in favor of natural gas-fired generation because of environmental concerns and the increased availability and low cost of natural gas and the ability of gas-fired generation to respond quickly to the variability of renewable generation.  Although gas pipelines have the ability store gas in the form of increased pressure in the piepline, i.e. linepack, this storage is limited.  In the future, linepack will be increasingly exercised as more gas-fired generation is used to balance increasing amounts of wind generation.  Larger swings in gas pipeline pressure (linepack) affect the ability of the pipeline to deliver gas to the generators, creating reliability implications that cascaded across these two infrastructures.

% The degree of coupling has increased in the U.S. for a number reasons. First, the development of fracking technology \cite{fracking} along with an environmental push to retire coal plants led to impressive growth in the gas-fired generation. Second, gas generators have the capability to respond quickly, contribute to frequency control (primary and secondary), and respond to renewable fluctuations. Such capabilities have introduced fast fluctuations in gas consumption that are correlated across the spatial extent of a the power system. Third, even though the gas system does not require instantaneous mass balance, acceptable boundaries for the internal (line pack) reserve of the system are limited.

In this initial work, we have assessed the impact of fluctuating consumption by gas-generators on pipeline pressure. We start by splitting the gas flow equations for pipelines into two parts.  The first is a stationary part that is time-independent and reflects the gas flows scheduled by the gas markets and gas compressor deployment determined by the pipeline operator.  The second is a representation of the fluctuations around the scheduled flows created by linearizing the gas flow equations about the scheduled flows and compressor operations.  From this linearized model, we can predict the probability that a set of stochastic gas loads will cause the pipeline pressure to violate an engineering or contractual pressure limit and create a reliability concern for the pipeline operator or the electrical grid operator.  By making assumptions about the nature of the gas consumption fluctuations, this probability can be expressed in an algebraic form that is convenient for integration into a gas flow/gas compressor optimization problem where the probability can be a constraint or part of the objective to limit the likelihood of a pipeline reliability issue.

We applied the theoretic results to a realistic model based on the Transco pipeline. Our computational experiments with the Transco model revealed the following interesting observations.  First, the probability of large pressure fluctuations is highest at locations in the pipeline where the gas flow experiences a reversal  (in and around the New York/New Jersey area for the Transco pipeline).  Second, increasing the stress on the pipeline by increasing gas flow rates leads to higher probabilities of large pressure fluctuations.  Third, rearranging pipeline flows, e.g. by increasing purchases from the Marcellus Shale at the expense of gas from the Gulf, can decrease the probability of large pressure fluctuations by moving the gas sources closer to the gas loads.

The results of this paper suggest a number of interesting directions for future research.
\begin{itemize}
\item The linearization and asymptotic assumptions described here need to be validated against direct dynamic (transient) simulations of gas flows in variety of situations. Most existing work on such validations \cite{84Osi,87TT,98TT,00ZA,11DF,12ABG} uses single pipe models.  The challenge is to develop fast computational algorithms for transient problems with  mixed (initial and boundary) conditions over large and loopy gas networks.
\item Our dynamic method applies to gas networks with loops, back flows, bi-directional compression and other complications. We will extend the experimental study to other current and planned networks in the U.S. and Europe.
\item Extending the probabilistic risk framework to the complications mentioned above requires extending the methods of \cite{13MFBBCP} to create efficient optimization algorithms for gas networks with loops.
\item
Compressor positions are assumed fixed by the OGF and greedy solution
methods presented here.  However, compressor position has a significant effect
on pressure fluctuations in its vicinity.  A future direction will be to
formulate a compressor dispatch scheme and use it to analyze the
effect of varying compressor position on pressure fluctuations near
the compressors.

\item Incorporation of the probabilistic risk measures into OGF formulations to directly account for this risk.  A promising direction is the chance constrained methodology developed in \cite{14BCH}.
\item This work suggests a new mathematical, statistical and computational foundations necessary to address the comprehensive strategic problems of re-organizing the existing system of energy trading (in the U.S. and elsewhere).  Such a reorganization is required to reduce inefficiencies in how power and gas markets interact \cite{2010MITEI,2013MITEI,14TA}.
\item It is not realistic to expect (at least not in US) that gas and power markets will merge in the near future. However, it is important to account for effects of mutual dependencies. In particular, incorporating effects of gas pressure fluctuations and uncertainty into planning and operations of power systems with significant penetrations of renewables and with gas turbines involved in balancing the renewable fluctuations is a very promising future direction for research. On the other hand it is as important to account for the effect of ramps in gas consumptions at generators on the gas flow optimization.
\end{itemize}

% %{\color{red} !!!!! Acknowledgments are needed only for the final submission - hide before submitting}
%\section{Acknowledgments}

%\input{Acknowledge.tex}

{\small
\bibliographystyle{IEEEtran}
%\bibliography{../Bib/GasFlow,../Bib/Russian}
%\bibliography{GasFlow,Russian,RefConrado,GasFlowSM}
\bibliography{GasNew,Russian,RefConrado,GasFlow,GasFlowSM}
}

\end{document}